\documentclass[11pt,twoside]{article}
\usepackage{CAGN2016}
\usepackage{graphicx}

\usepackage[T1]{fontenc} 

\usepackage{latexsym}
\usepackage{verbatim}

\begin{document}

\vskip 1.0cm
\markboth{O.~Cavichia et al.}{Improving the determination of chemical abundances in planetary nebulae}
\pagestyle{myheadings}
%
%
\vspace*{0.5cm}
\parindent 0pt{Contributed  Paper}


\vspace*{0.5cm}
\title{Improving the determination of chemical abundances in planetary nebulae}

\author{O.~Cavichia$^1$, R.D.D.~Costa$^2$, W.J.~Maciel$^2$ and M.~Moll\'a$^3$}
\affil{$^{1}$Instituto de F\'isica e Qu\'imica, Universidade Federal de Itajub\'a, Av. BPS, 1303, 37500-903, Itajub\'a-MG, Brazil\\
$^{2}$Instituto de Astronomia, Geof\'isica e Ci\^encias Atmosf\'ericas, Universidade de S\~ao Paulo, 05508-900, S\~ao Paulo-SP, Brazil\\
$^{3}$Departamento de Investigaci\'on B\'asica, CIEMAT, Avda. Complutense 40, E-28040 Madrid, Spain}

\begin{abstract}

Planetary nebulae are the products of the evolution of low and intermediate mass stars. The chemical property studies of these objects give important information about the elemental abundances as He, O, Ne, Ar, S and  their modifications associated with the evolution of the progenitor stars. The determination of accurate abundances in planetary nebulae is important from the perspective of the stellar evolution as well as the formation and chemical evolution of galaxies. Recently, new HeI emissivities and ionization correction factors (ICFs) were published in the literature. In this work, these new parameters are used in a code for the determination of chemical abundances in photoionized nebulae. This code is used for the recompilation of the chemical abundances of planetary nebulae from the Galactic bulge observed previously by our group and also for the determination of new chemical abundances of a sample of planetary nebulae located near the Galactic centre.  The new emissivities and ICFs slightly modified the elemental abundances of He, N, O, Ar and Ne. On the other hand, S abundances are higher than previous determinations. The new ICFs can contribute to solve partially the sulphur anomaly.
\end{abstract}

\section{Introduction}

The determination of accurate chemical abundances in planetary nebulae (PNe) is very important,  since these objects can contribute to studies regarding stellar evolution as well as the chemical evolution of galaxies. Recently, new He\,{\sc i} emissivities have become available in the literature through the work of  Porter et al. (2012) and subsequently corrected by Porter et al. (2013), hereafter PFSD13. These are the most recent He\,{\sc i} emissivities, and collisional effects are already included in the emissivities calculation. In this work the emissivities of PFSD13 are adopted in order to calculate the He\,{\sc i} abundances and provide accurate He\,{\sc i} abundances. Usually, the spectral range of the observations is not sufficient to observe all the necessary lines of a given ion, and it is not possible to calculate the total abundance of a particular element by the direct sum of the ionic abundances of all the ions present in a nebula. Instead, it must be calculated by means of the ionization correction factors (ICFs). One of most frequently used ICFs in the literature are those from Kingsburgh \& Barlow (1994), hereafter KB94. However, recently Delgado-Inglada et al. (2014), hereafter DMS14, have published new ICFs formulae based on a grid of photoionization models and they have computed analytical expressions for the ICFs of He, O, N, Ne, S, Ar, Cl and C. 
These new ICFs incorporate more recent physics and  are derived using a wider range of parameters of photoionization models than the previous ICFs from KB94. Another advantage of the new ICFs provided by DMS14  is the possibility to compute the errors in the elemental abundances introduced by the adopted ICF approximation. They provide analytical formulae to estimate error bars associated with the ICFs, something not possible until their work. According to  DMS14, the oxygen abundances are not expected to be very different from those calculated with the ICFs of  KB94. On the other hand, the abundances of N, S, Ar, Ne calculated with the new ICFs show significant differences. However, a direct comparison between the abundances of N, S, Ar and Ne calculated with both ICFs is missing in DMS14. The aim of this work is to use these ICFs and new emissivities to derive chemical abundances of a PNe sample located in the Galactic bulge and also near the Galactic centre (GC).  

\section{The sample and abundance determinations}

The new emissivities and ICFs were used to recalculate the elemental abundances of PNe located in the Galactic bulge from Cavichia et al. (2010), hereafter CCM10. We followed the same procedures described in CCM10, nonetheless using the new emissivities from PFSD13 and ICFs from DMS14. The H$^+$ emissivities are from Aver et al. (2010) and He\,{\sc ii} emissivities are from Osterbrock \& Ferland (2006). Also, In 2009 we started an observational program aimed at carrying out a spectroscopic follow-up of high-extinction GBPNe located within 2 degrees of the GC from the catalogue of Jacoby \& Van de Steene (2004). The Goodman spectrograph at the 4.1 m SOAR telescope at Cerro Pach\'on (Chile) was used to perform spectroscopic follow-up of 33 objects located within 2 degrees of the GC, in a region of a very high-level of reddening. From those objects, 15 had spectra with acceptable quality to derive physical parameters and chemical abundances. The optical spectra of the GBPNe near the GC suffer for high-level of extinction caused by the material near the Galactic plane and also in the central regions of the Galaxy. As a result, important diagnostic lines as [O\,{\sc iii}] $\lambda$4363 \AA \ and [N\,{\sc ii}] $\lambda$5755 \AA \ do not have enough S/N ratio to obtain the electron temperature from the temperature diagnostic diagrams. Other important temperature-sensitive lines are those from S$^{+ 3}$. In order to observe the NIR [S\,{\sc iii}] lines, we started an observational program in 2012 at the Observat\'orio Pico dos Dias (OPD) of  National Laboratory for Astrophysics (LNA, Brazil) with the 1.6 m Perkin-Elmer telescope and a Cassegrain Boller \& Chivens spectrograph was used. The details of the observations, data reduction, abundances determinations and results can be seen in Cavichia et al. (2017), hereafter C17.

\section{Results}

\begin{figure}[!ht]
\centering
\includegraphics[width=6cm]{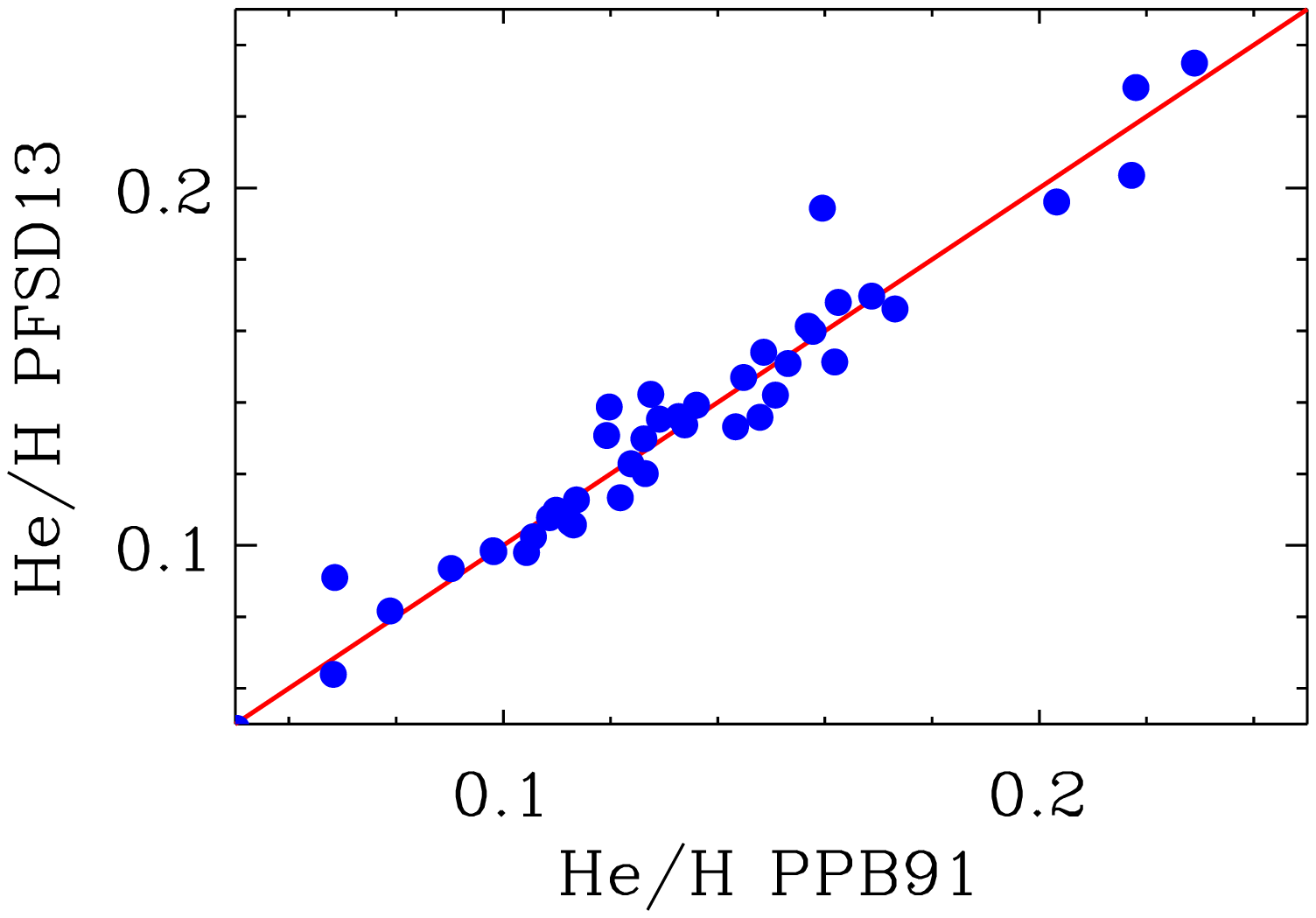}
\includegraphics[width=6.5cm]{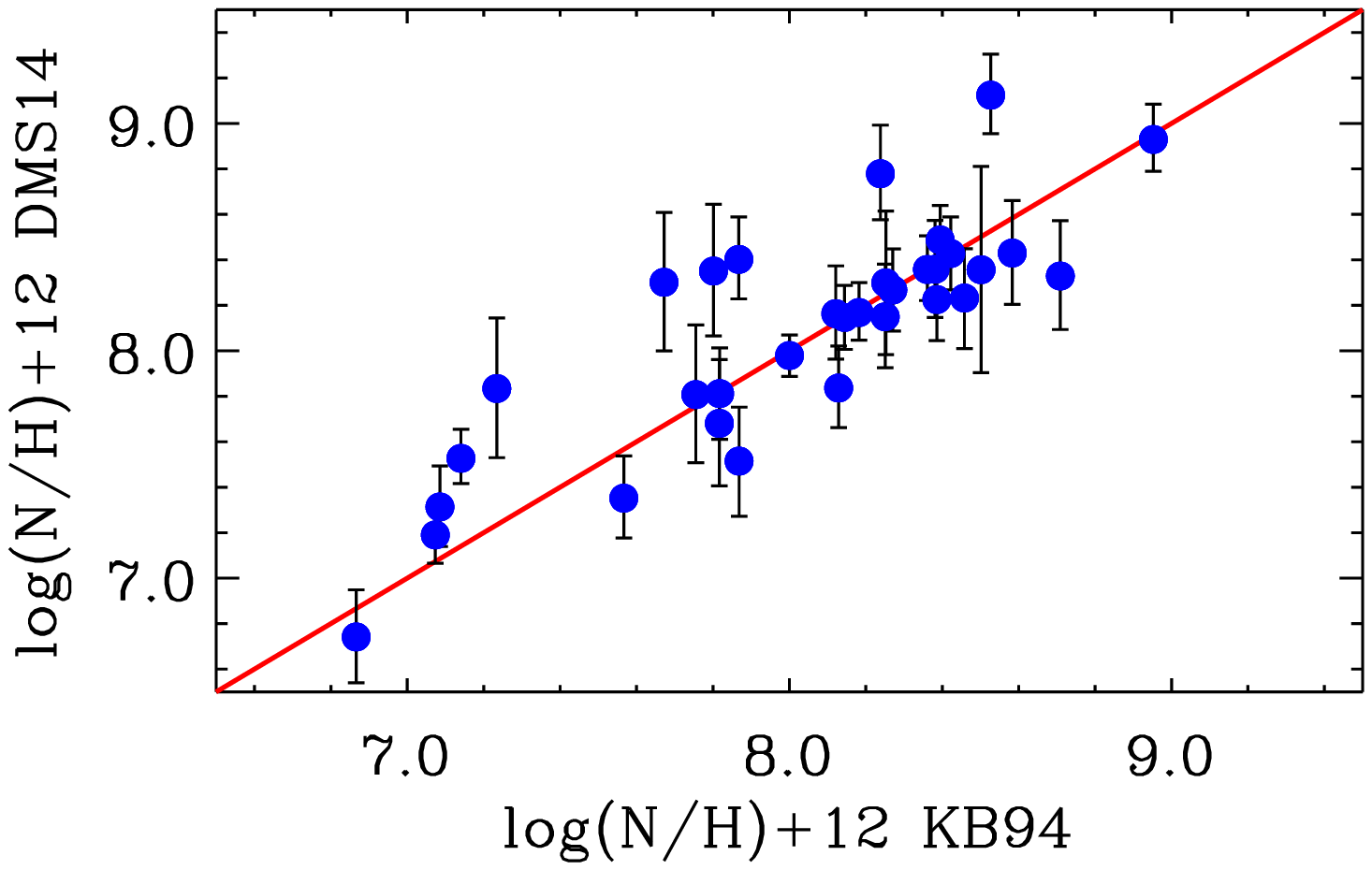}
\caption{Left: Comparison between He/H abundances from CCM10 (outer bulge PNe)  calculated using the emissivities from Porter et al. (2013, PFSD13) and Pequignot et al. (1991, PPB91). Error bars include only the errors in the ICFs. The red continuous line is the one to one relation. Right: The same as left panel but for $\log(\mbox{N/H})+12$.\label{fig1}}
\end{figure}

The abundances derived using the new He\,{\sc i} emissivities from PFSD13 are compared in Fig. \ref{fig1} (left panel) with those from Pequignot et al. (1991) hereafter PPB91. The PNe sample used for the comparison is that one from CCM10. As can be seen, some small differences arise when using the new emissivities. From now on, the emissivities from PFSD13 will be used to analyze the differences in the ICFs. A direct comparison between the ICFs from DMS14 and KB94 is provided  for N in Fig. 1 (right panel). Most of the abundances are compatible, however for a few objects the N abundance is higher than the one obtained by using the previous ICF. This is regarded as the new ICFs formulae use a wider range of parameters of photoionization models than the previous ICFs .

\begin{figure}[!ht]
\centering
\begin{minipage}{0.9\textwidth}
\includegraphics[width=6cm]{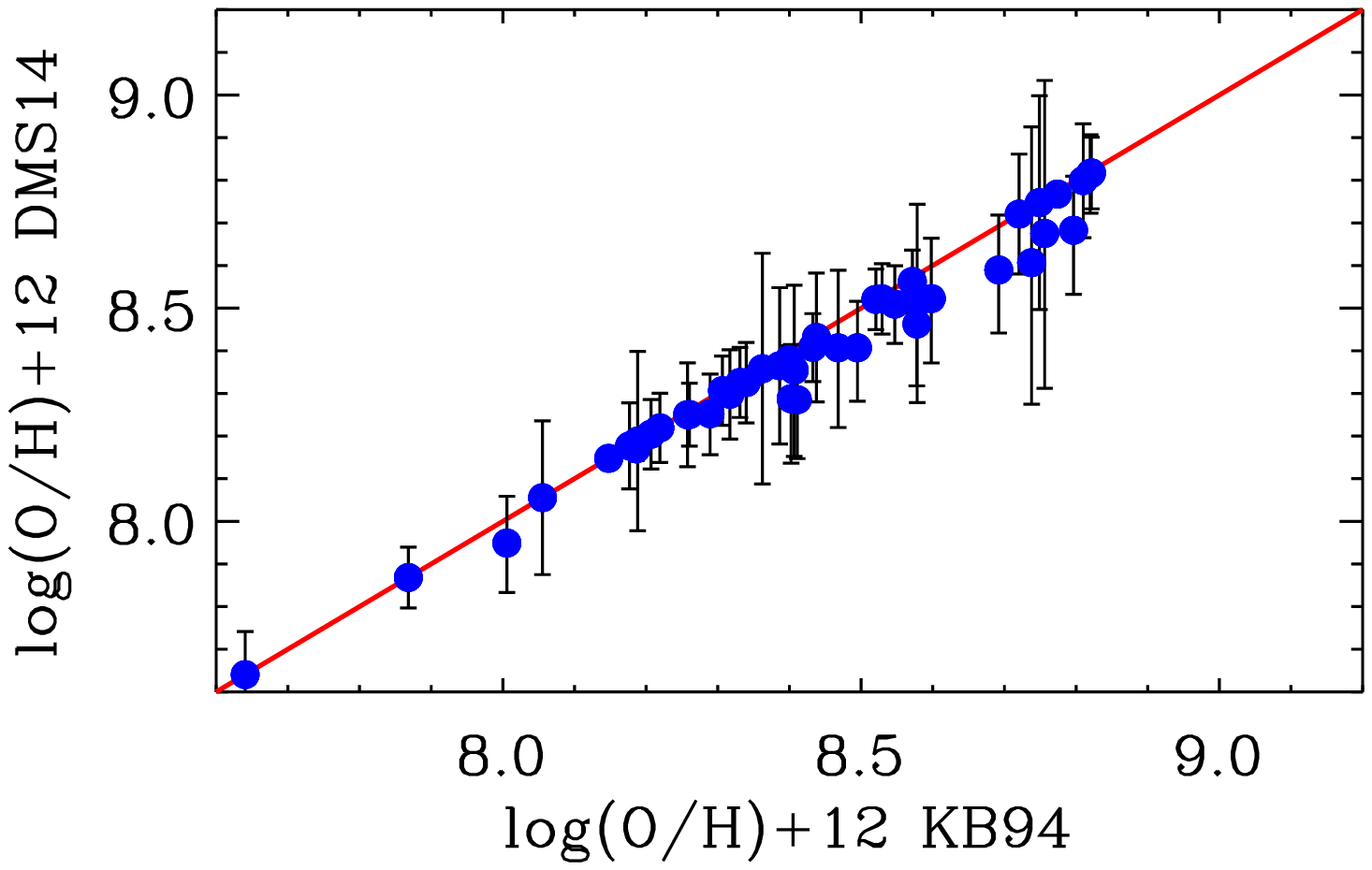}
\includegraphics[width=6cm]{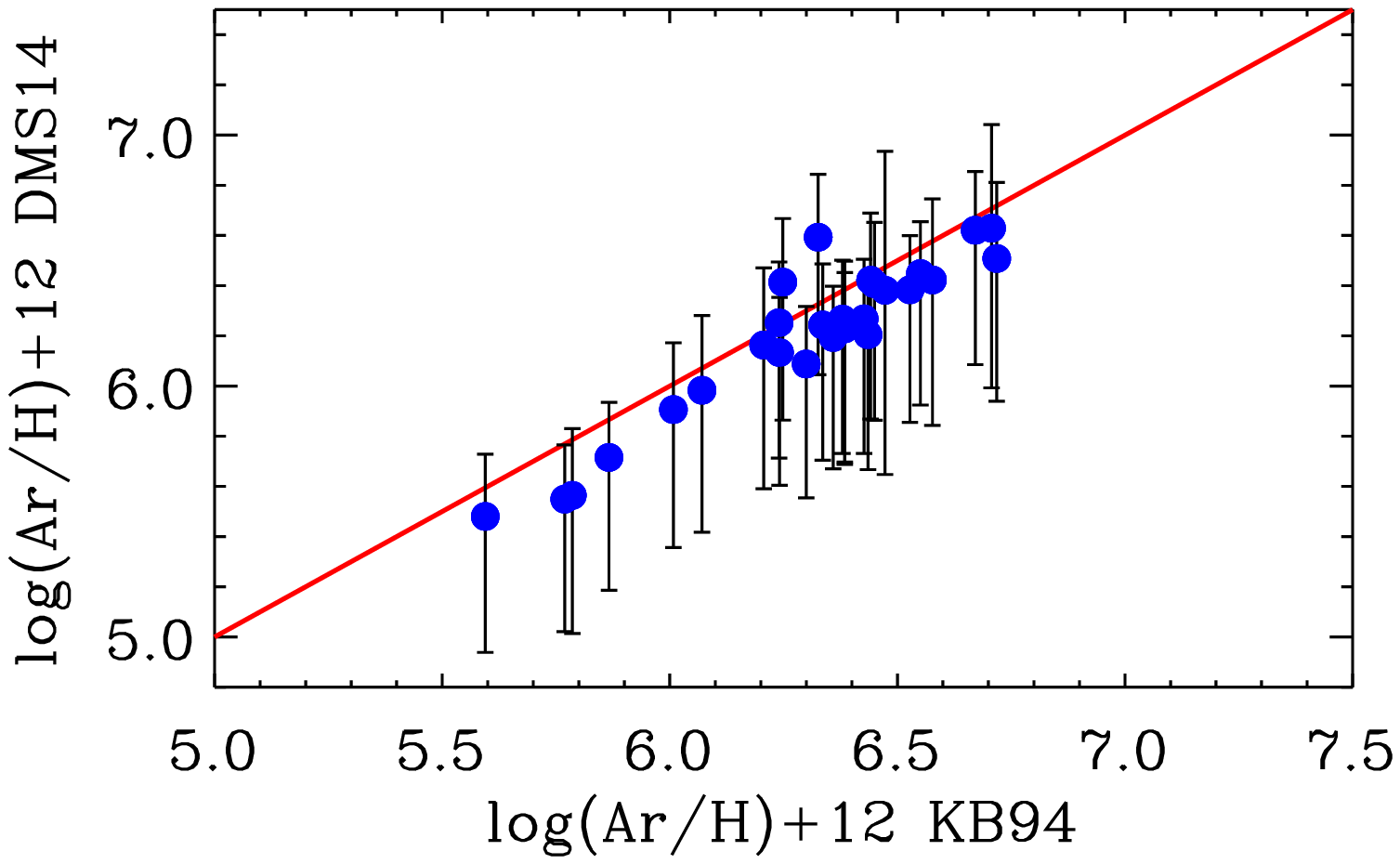}
\includegraphics[width=6cm]{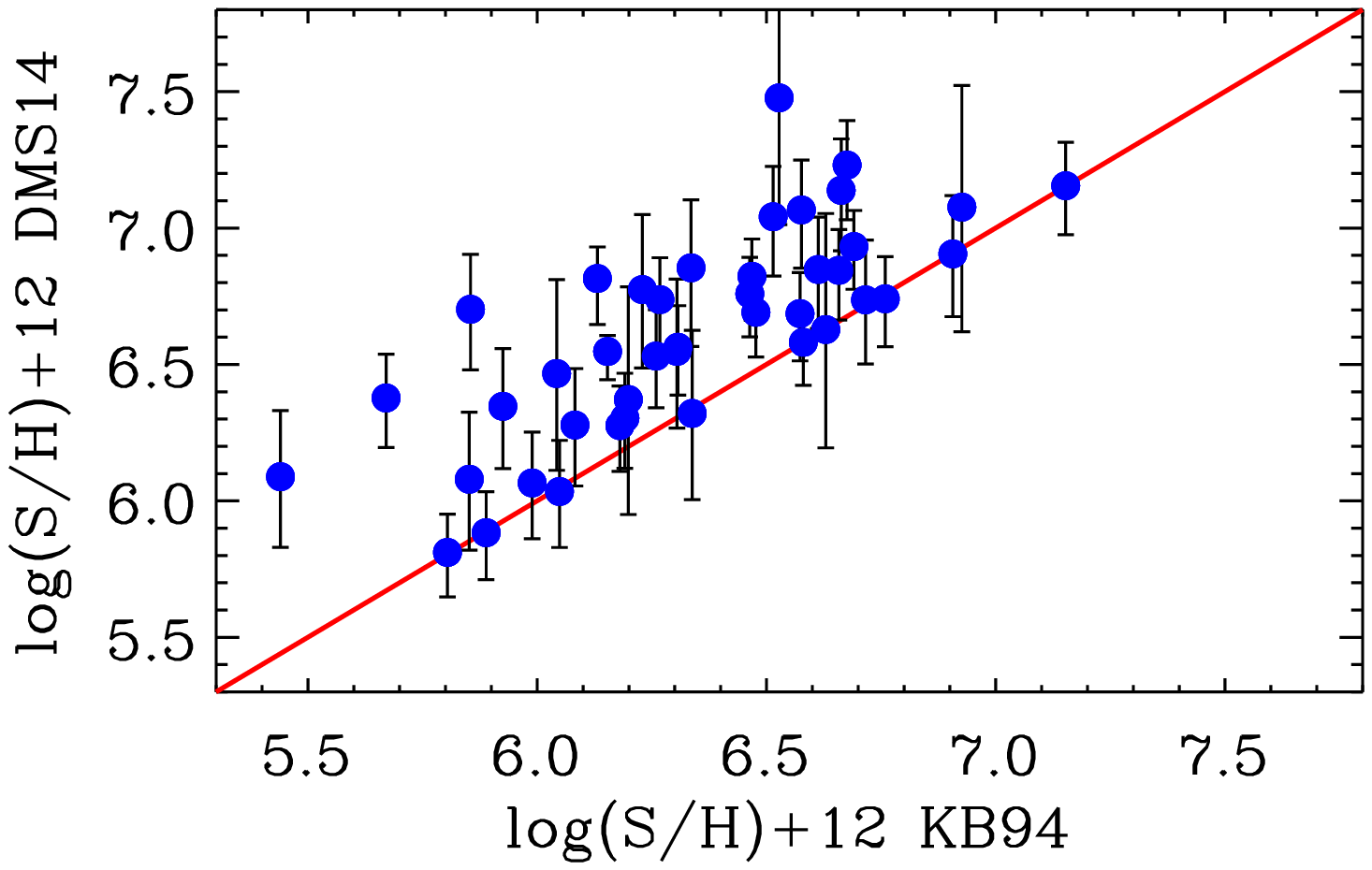}
\includegraphics[width=6cm]{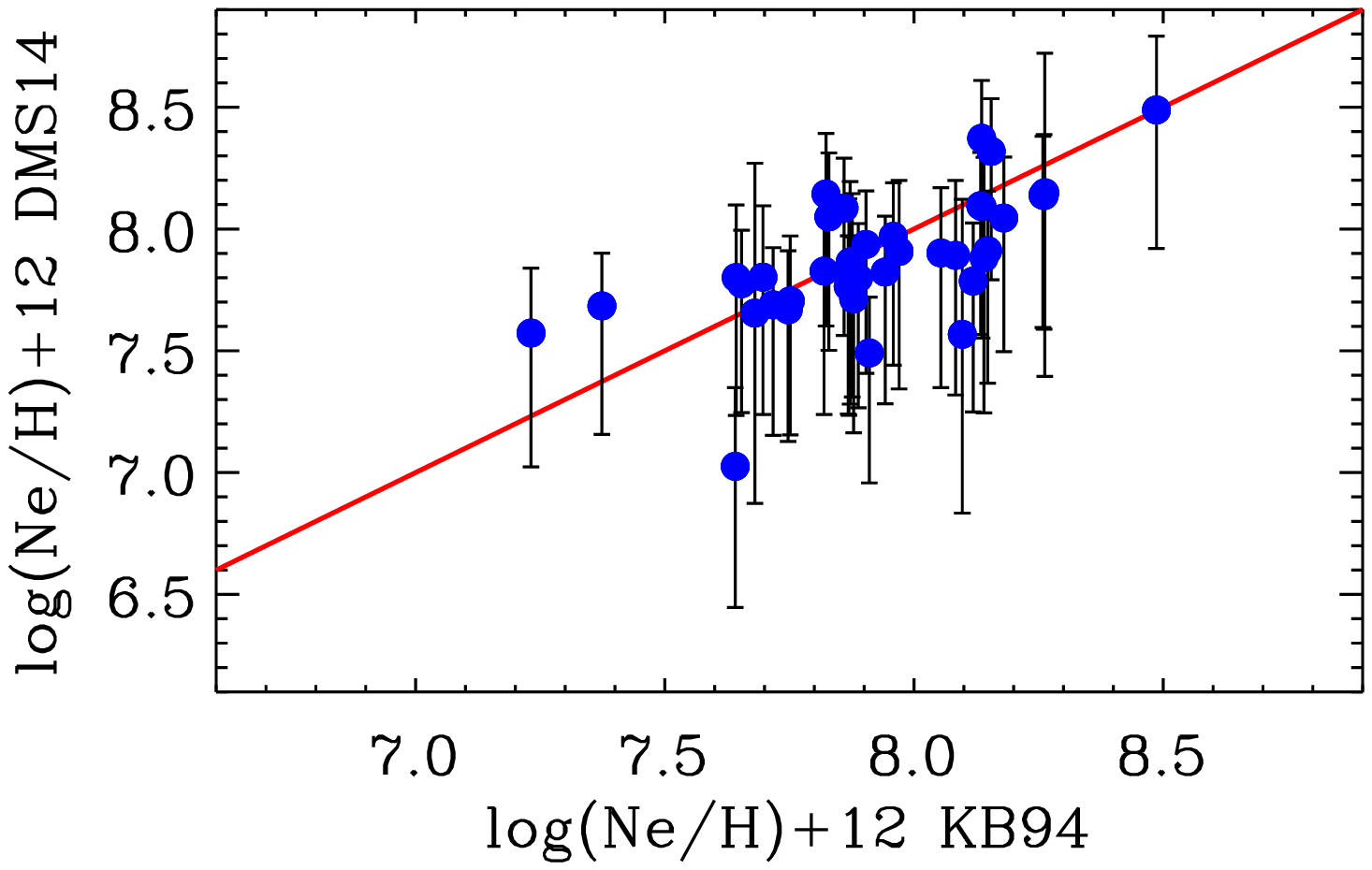}
\end{minipage}
\caption{Left: The same as Fig. \ref{fig1} for $\log(\mbox{O/H})+12$ (top left), $\log(\mbox{Ar/H})+12$ (top right), $\log(\mbox{S/H})+12$ (bottom left) and $\log(\mbox{Ne/H})+12$ (bottom right).\label{fig2}}
\end{figure}

 In Fig. \ref{fig2} the same confrontation is done for O, Ar, S, and Ne. Small differences are noted for O, Ar and Ne for both ICFs. However, In the case of S the new ICFs provide higher abundances than the previous ones from KB94. The new ICFs from DMS14 can, at least partially, solve the ``sulphur anomaly problem'' in PNe. 

\begin{figure}[!ht]
\centering
\begin{minipage}{1.0\textwidth}
\centering
\includegraphics[width=6cm]{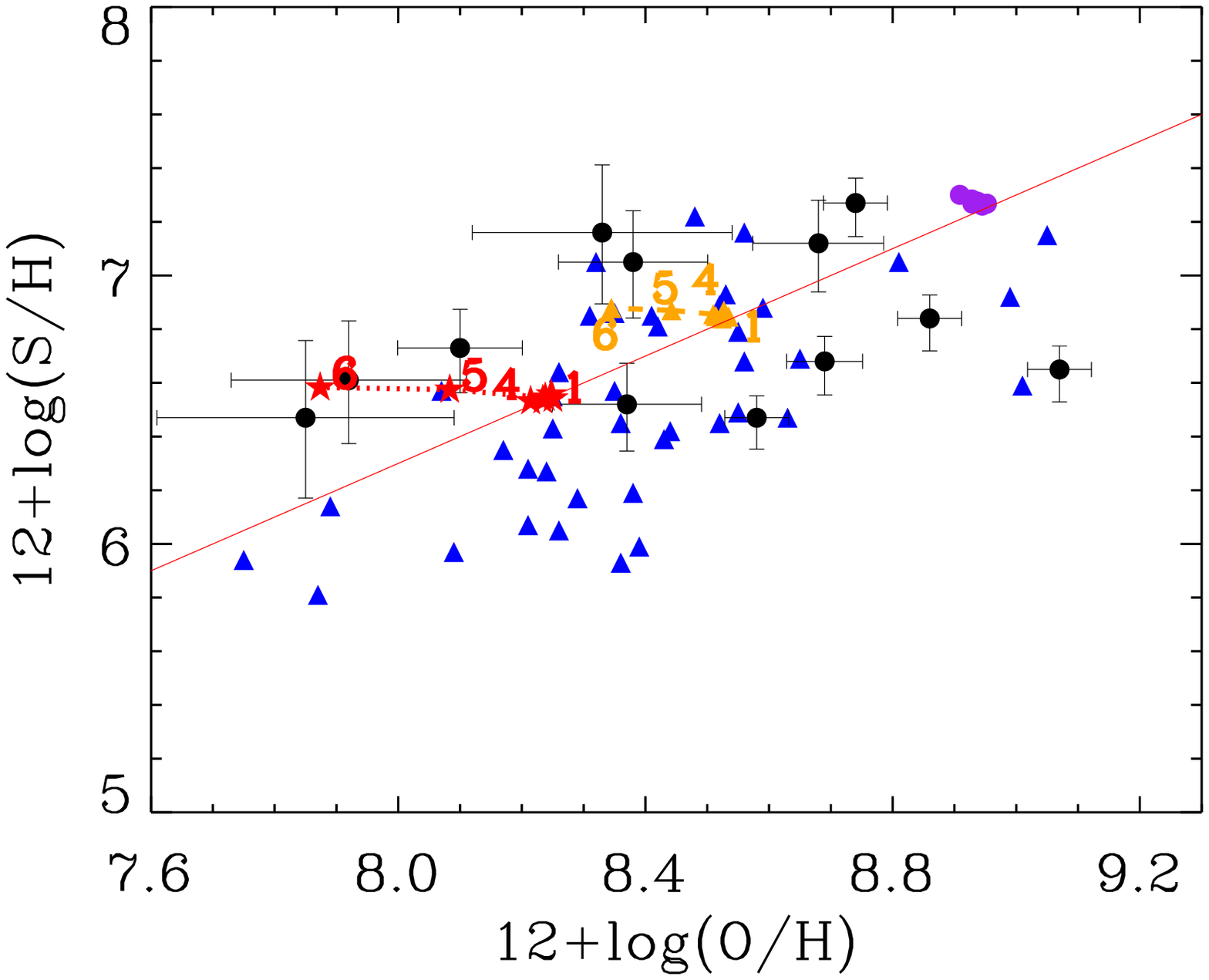}
\includegraphics[width=6cm]{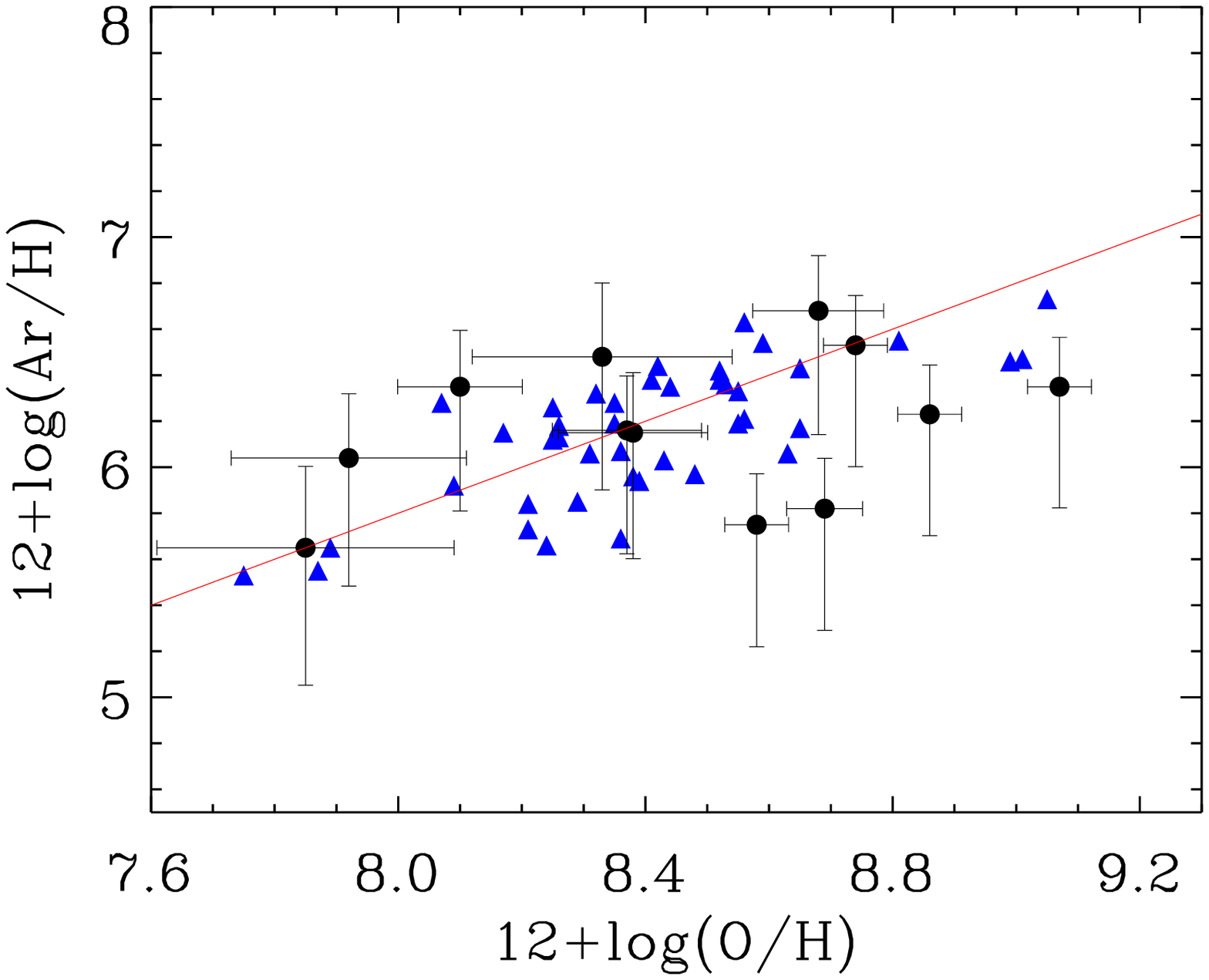}
\end{minipage}
\caption{Left: $12+\log(\mbox{S/H})$ vs. $12+\log(\mbox{O/H})$; Right: $12+\log(\mbox{Ar/H})$ vs. $12+\log(\mbox{O/H})$.  Filled circles with error bars are data from C17 (PNe near the GC), while filled blue triangles are the data from CCM10 (outer bulge PNe). The symbols with numbers represent the results of the models from Karakas (2010) for a given value of $Z$: red stars joined by dotted line for $Z=0.004$; yellow triangles joined by dashed line for $Z=0.008$;  purple circles joined by dash-dotted line for $Z=0.02$.  The numbers give the initial masses of the individual models in M$_{\sun}$ units. The red continuous lines are straight lines with slopes equal to one.\label{fig3}}
\end{figure}

The behaviour of Ar/H vs. O/H and S/H vs. O/H are presented Fig. \ref{fig3}.  In this figure both samples from CCM10 (outer bulge PNe) and C17 (PNe near the GC) are included. A correlation is found for these elemental abundances for both samples, although S/H vs. O/H for CCM10 data the correlation is not as tight as in the case of our data. S abundances of C17 were determined using both S$^+$ and S$^{++}$ ions, contributing to reduce the uncertainties in the S abundances. The predictions of evolution models by Karakas (2010) at Z = 0.004, 0.008 and 0.02 are displayed in the graph for S/H vs. O/H. Clearly, models show that S is not modified by stellar nucleosynthesis, independently of the initial stellar mass. On the contrary, O is expected to be modified in progenitor stars heavier than 4~M$_{\sun}$ at low-metallicity environments.  In the case of Ar, the models by Karakas (2010) do not give predictions for these elements, so that the abundances could not be compared with theoretical results. 

\begin{figure}[!ht]
\centering
\includegraphics[width=8cm]{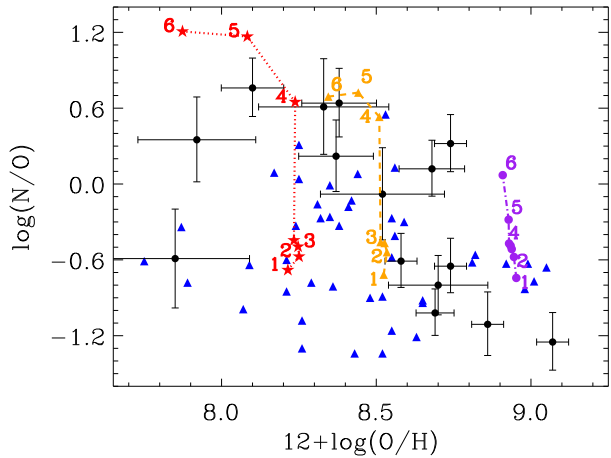}
\caption{N/O ratio as a function of oxygen abundances. Symbols are as in Fig. \ref{fig3}.\label{fig4}}
\end{figure}

Fig. \ref{fig4} shows the N/O ratio as a function of O abundances. We can observe a fair agreement between the data and the models by Karakas (2010) for models $Z=0.008$ and $0.02$. An important difference between the samples of C17 and CCM10 is observed in this figure: in CCM10 some points are compatible with the lower metallicity model ($Z=0.004$) by Karakas (2010) and also lower initial masses ($< 4 \mbox{M}_{\sun}$). In the sample of C17 (PNe near the GC)  the majority of PNe have abundances compatible with models at higher metallicities. This can indicate a faster chemical enrichment taking place at the GC, compared with the outer regions of the Galactic bulge. However, this result should be interpreted with some caution, since due to the high interstellar extinction in the direction of the GC it is very difficult to define a metallicity-unbiased sample.

\section{Conclusions}

We have implemented the new ICFs from Delgado-Inglada (2014, DMS14) for the elemental abundances determination and also the new He\,{\sc i} emissivities from Porter et al. (2013, PFSD13). The abundances of Cavichia et al. (2010, CCM10) were recalculated with the new emissivities and ICFs. A direct comparison between the He abundances using the new He\,{\sc i} emissivities and those from Pequignot et al. (1991, PPB91) resulted in small differences. The new ICFs from DMS14 were compared with those from Kingsburgh \& Barlow (1994, KB94). The results show that N, O, Ar and Ne abundances are compatible within the uncertainties. However, S abundances derived with the new ICFs are higher by 0.2 dex in average. The new ICFs from DMS14 can, at least partially, solve the ``sulphur anomaly problem'' in PNe. 

We have performed spectrophotometric observations with the 4.1 m SOAR (Chile) and the 1.6 m OPD/LNA (Brazil) telescopes to obtain physical parameters and chemical abundances for a sample of 15 planetary nebulae located within 2 degrees of the galactic centre (see details in Cavichia et al. 2017). S abundances were derived using optical and NIR lines, reducing the uncertainties associated with S ICFs. The abundances predicted by Karakas (2010) for stars of different initial masses and metallicities were used to constrain the masses and initial metallicity of the progenitor stars. An important difference between the sample located near the Galactic centre and PNe located in the outer parts of the bulge is observed. In our previous work (CCM10) some points are compatible with the lower metallicity model ($Z=0.004$) by Karakas (2010) and also lower initial masses ($< 4 \mbox{M}_{\sun}$). In the PNe located near the Galactic centre, the majority of PNe have abundances compatible with models at higher metallicities. This can indicate a faster chemical enrichment taking place at the Galactic centre, compared with the outer regions of the Galactic bulge. 

\acknowledgments This work was partially supported  by CAPES, FAPESP and CNPq.

\end{document}